# The Roles of Astronomers in the Astronomy Education Ecosystem: A Research-Based Perspective

Stephen M. Pompea[1] and Pedro Russo[2]

1      National Optical Astronomy Observatory, Tucson, Arizona, U.S.A.; email: spompea@noao.edu
2      Astronomy and Society Group, Leiden Observatory and Department of Science Communication and Society, Leiden University, the Netherlands; email: russo@strw.leidenuniv.nl

## Abstract

Astronomers have played many roles in their engagement with the larger astronomy education ecosystem. Their activities have served both the formal and informal education communities worldwide, with levels of involvement from the occasional participant to the full-time professional. We discuss these many diverse roles, giving background, context, and perspective on their value in encouraging and improving astronomy education. This review covers the large amounts of new research on best practices for diverse learning environments. For the formal education learning environment, we cover pre-university roles and engagement activities. This evidence-based perspective can support astronomers in contributing to the broad astronomy education ecosystem in more productive and efficient ways and in identifying new niches and approaches for developing the science capital necessary for a science literate society and for greater involvement of underrepresented groups in the science enterprise.

## Summary points

— Astronomy Education is a vast and well-established field, with much relevant research for the practitioner.

— Astronomers can contribute in a myriad of ways to the educational ecosystem; however, a basic understanding of learning theory, education culture, and best practices is a requirement.

— Astronomy can and should contribute to a more literate society and also to a more equal and inclusive society, by engaging students from diverse underprivileged, underrepresented, and under-reached communities.

— Astronomers should approach education programs in partnership with well-established education stakeholders and base their program on evidence-based approaches.

— Contributions to education by professional astronomers need to be properly recognised and rewarded throughout an astronomer's career.

— Astronomy education needs long-term institutional support and commitment, including structural funding for people and projects.

## 1.  INTRODUCTION

The universe is larger and more diverse, dynamic, and enigmatic than our ancestors could have imagined when they first gazed at the stars. The captivating and mysterious nature of the universe makes astronomy fascinating to the public and a fertile ground for the imagination of young and old minds alike. Astronomers have taken the responsibility to engage with the public, with educational commitments representing a large portion of the well-established education and public engagement (EPE) subfield of astronomy. The term "outreach", which refers to one-way communication approaches, has gone out of favor since it has associations that are less intentional and collaborative. Since the 1960s, mainly due to NASA's Apollo program (Scott & Jurek 2014), institutional – and individually initiated EPE have been part of the astronomy enterprise. A 2004 survey found that 58% of scientists engage in some form of outreach (National Science Board 2004). According to a 2015 survey of 2,580 members of the IAU, 87% of the astronomers in the world engage in EPE activities of some type (Entradas & Bauer 2018).

The EPE field has been categorized by Morrow (2000) as well as Christensen and Russo (2007) as a continuum of overlapping subfields, from formal education programs to policymaker and public relations support (Fig 1). Each subfield has its own audiences, methodologies, approaches, products, and evaluation methods. In this article, we will focus only on pre-tertiary (or pre-university) formal and informal education, ignoring media relations and the many other forms of science communication. We also will not cover the rich field of university astronomy education, which has some of the same themes presented here.

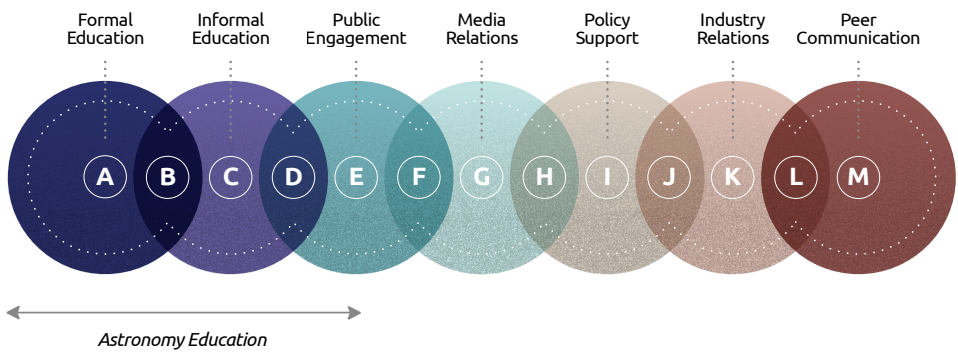

Figure 1

Educational activities occupy the left part of the spectrum and serve the broad formal and informal educational ecosystem. Each circle of influence supports multiple audiences and stakeholders and utilizes a variety of engagement activities and methodologies to achieve its goals. Based on: Morrow (2000) and Christensen & Russo (2007)

Some examples of engagement activities illustrated in **Figure 1** are:

— **AB Formal Education:** Writing high school textbooks, developing instructional materials, creating bibliographies for teaching, translating educational materials, creating simulations for use in schools, review of materials, conducting teacher professional development, classroom visits, demonstrations at schools, loaning resources to schools, teaching coding, refurbishing a science lab, bringing teachers and students into research groups, serving on school boards

— **BCD Informal Education:** using robotic telescopes, mentoring to astronomy clubs, museum exhibit development, developing planetarium shows, demonstrations at science centers, conducting professional development for informal educators, helping a children's museums, support of touristic observatories, organizing citizen science projects, working with amateur clubs, doing sidewalk astronomy observing, observatory tours, giving a science talk in a pub, participating in teen cafes, observing events and star parties, serving on museum advisory boards

— **DEFGHIJKM Public Engagement, Communication, et al.** YouTube videos, creation of standalone simulations, TV/Radio, Podcasts, printed-media, creating and curating astronomical images, creating press releases, holding press conferences, media courses for scientists, public hearings

Simply participating in these various activities in not enough. In all of these activities, scientists should be aware of their motivation for participating and of the motivation of the participants in order to create intentional, meaningful interactions, which move beyond the deficit/expert model. Authentic engagement activities in any of these categories must have a depth of engagement (adequate time), a balance of voice (a dialogue not dominated by the scientist), and a balance of power and control (Storksdieck et al. 2016). For example, a scientist giving a lecture may be performing a valuable service, but it may not be a true engagement activity as it often lacks the opportunity for significant dialogue. It is largely one-sided in communications flow (the astronomer has the answers and the audience needs to learn) and in power/control. However, the scientist can take steps to improve the level of engagement by increasing the time for dialogue and questions, meeting informally before or after the presentation, making the presentation available, and taking steps to decrease the perception of a power differential

Astronomers at all career stages put astronomy education as one of their top activities in engaging with society. Almost a quarter of senior professional astronomers report conducting activities with schools (Entradas & Bauer 2018). This article provides a timely review of significant developments in astronomy education, establishing connections across education, science education, physics education, and, specifically, the broad field of astronomy education research. This review is aimed at interested astronomers who want to expand their knowledge of science or astronomy education. It can be valuable to both students who want to organize a school activity or more senior astronomers who want to develop a more comprehensive education programs to complement their research activities.

Our goal is two-fold: to familiarize this vast network of practitioners with advances in the field in order to help them work more efficiently and effectively, and to help them avoid the frustration that comes from misguided approaches and novice-level misunderstandings.

## 2. THE ROLE OF ASTRONOMY IN EDUCATION

Several education researchers have demonstrated the relevance of astronomy for education, both in terms of students' interests as well as cognitive development. Sjøberg and Schreiner (2010) have shown that space and life in outer space are the most interesting science topics for learners (both boys and girls) in a survey of over 30 countries. Liberman (2012) also shows that learning about astronomical objects promotes abstract thought. People tend to visualize proximal objects using concrete and detailed images, while visualizing distant objects—such as those common in astronomy—in a more abstract and decontextualized way. Liberman found that after spatial distance priming, learners are more creative in both fluency and originality. Furthermore, astronomy is a very comprehensive discipline encompassing a broad range of formal education subjects (Figure 2) including biology, geology, chemistry, physics, engineering, philosophy, and history. This makes astronomy a natural umbrella science to a more holistic education.

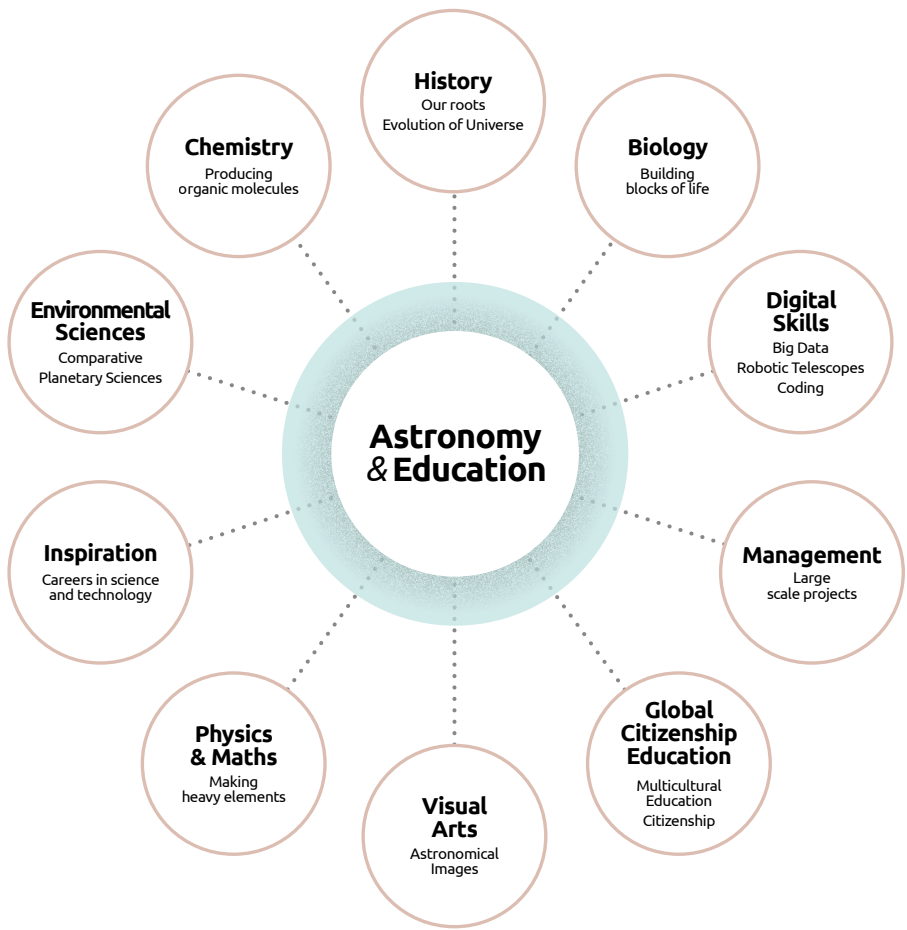

Figure 2

Connections of astronomy with other formal education subjects including biology, geology, chemistry, physics, environmental science, engineering, philosophy, and history. Based on G. Miley/IAU (2009).

## 2.1    Global Roles

The well-established Programme for International Student Assessment (PISA) was intended to evaluate educational systems by measuring the scholastic performance of 15-year-old students. Their new Global Competence Framework, created in 2018, measures student competence in understanding the wider world around them (OECD, 2018). For example, examining a global issue, such as climate change, requires knowledge of that issue, the skills to transform this awareness into a deeper understanding, and the attitudes and values to reflect on the issue from multiple cultural perspectives. Astronomy, with observational facilities that are cross-national and with its global research collaborations, can and should contribute greatly to the new Global Competence Framework. It is a pressing necessity, not a luxury, that students learn to examine global and intercultural issues.

What is the role of research astronomers in contributing to this enlarged global education role? Several policy documents (e.g. the US National Research Council, the US Astronomy and Astrophysics Survey Committee in 2001, the European ASTRONET Infrastructure Roadmap from 2008 and 2013) (Bode 2008; Robson 2014) have described the roles of the professional astronomy community in astronomy education as:

— To widely disseminate astronomical discoveries, and thus bring the excitement inherent in science to the broader public.

— To use the excitement that astronomy engenders to increase public understanding of science and scientific methods, and to make clear that science is a pathway to discovery, not just a collection of facts.

— To capitalize on the close involvement with astronomy, technology, and instrumentation to contribute to training the technical workforce.

— To contribute to a more equal and inclusive society, by engaging students from different underprivileged, underrepresented and under-reached communities (from women and girls to visually and otherwise impaired and indigenous and other ethnic minorities. To prepare future generations of preeminent astronomers who will contribute to a scientifically literate society.

These goals are highly relevant today, summarizing for the astronomy community the important roles that astronomers do play—and aspire to play—in participating in and improving science education. These roles are being served in a wide variety of ways. For example, astronomers around the world are serving as mentors to secondary students in the Einstein Schools Programme (IAU 2019) and as resources to teachers in long-standing programs such as Project ASTRO (Fraknoi et al. 1996). These are but two of the many roles that will be discussed in this review.

## 2.2    Barriers to an Expanded Role

There is evidence that astronomers would like to be more active in education (and public engagement) activities (Spuck et al. 2017), but are not because of barriers such as a lack of time and institutional support (Borrow & Russo, 2015). Another limiting factor is a lack of accessible, helpful research and pedagogical knowledge about science education, astronomy education, and education in general (Miller 2008). This review will contribute to increasing astronomers' knowledge about the foundations of science education research applied to astronomy (Section 3), activities in the school and classrooms (Section 4), activities in out-of-school settings (Section 5), and innovative approaches to astronomy education (Section 6).

Astronomers can and should play another important role in astronomy education, by promoting the processes and values of science including rational inquiry, the scientific method, global citizenship, and the paramount importance of evidence and evidence-based decision making. This review will enable astronomers to contribute to this vision by presenting and discussing best practices in promoting an understanding of the process and authentic practices of science.

## 2.3    Scope of this Review

Astronomy education can be viewed as a continuum, from preschool efforts to lifelong learning, with many different approaches, audiences, and roles for astronomers (e.g. Gouguenheim 1998; Pasachoff & Percy 2005; Pasachoff et al. 2008). This review will explore the many different roles that astronomers can play in the pre-university level educational community. While we focus on preschool, primary, and secondary schools, we realize that astronomy education at the community college (Fraknoi 2004) and university levels shares many of the same educational themes as secondary school education. There are innovations that make classrooms more interactive (e.g. Mazur 1999; Duncan 2006) coupled with new assessment and diagnostic tools (e.g. Hufnagel 2002) to help educators gain a deeper understanding of students' conceptual knowledge.

However, education at the undergraduate university level is a specialized field with a robust literature base for classroom practices and online learning (e.g. Bailey & Prather 2003; Prather et al. 2004; Slater 2008; Bailey & Lombardi 2015; Impey & Buxner 2019; Impey & Wenger 2019). Given the breadth and depth of the pre-university science education field, and the number of roles in which astronomers can contribute apart from that, we will not address university science education with the single brief exception of describing the relevant overlap with physics education research that applies to secondary school astronomy education.

## 3.    FOUNDATIONS OF ASTRONOMY EDUCATION

Even for the occasional practitioner, it is still critical to have a broad perspective of astronomy education research, especially in terms of how this field is entwined in the broader field of science education. Although the field of science education is only 80 years old (Abell and Lederman 2007), there is a significant body of research relevant to the practitioner, and recent trends in research may be particularly relevant, especially to an understanding of science learners and the effects of culture, gender, and society (National Academy of Sciences 2018). New relevant research on science teacher learning and the importance of discipline-specific teaching knowledge, termed pedagogical content knowledge (PCK), is also valuable. As noted physicist and science educator Carl Weiman asks so succinctly "Why not try a scientific approach to science education?" (Wieman 2007).

Our goal is to describe research that can aid in the improvement of astronomy teaching and learning all over the world, preferably in the same spirit as the excellent review by Abell and Lederman (2007). Science education research is an applied field and must address the concerns, needs, and questioners of practitioners in all varieties and venues of astronomy education. We hope that our selection and rendition of the key concepts will help bridge the gulf between research and practice and can suggest new areas where additional research will be valuable to the practitioner. Bybee and Morrow (1998) and Bybee (1998) have summarized the many roles that scientists can play in improving science education in diverse settings; hence, the need for a broad understanding of teaching and learning. Some relevant topics in science education research include inquiry as a guiding framework for science education, how best to teach the nature of science, contributions to systemic science education reform, how the science teacher learns, and their attitudes, beliefs, and state of knowledge (Abell & Lederman 2007; Michaels et al. 2008).

### 3.1    Science Literacy Development

Much of the research on science learning emphasizes the development of science literacy, which includes discipline-specific knowledge and practices, the values of science, and an understanding of the scientific process (Bybee 1997). Although improvement has been noted, it is generally agreed that most students and adults do not achieve even moderate levels of science literacy. Furthermore, the levels achieved are highly related to race, ethnicity, and social class (Anderson 2013; Blank & Langesen 2001). There is also a considerable body of research on quantitative or numerical literacy, which unfortunately, is lacking in students of all ages. Follette and McCarthy (2012) and Follette et al. (2017) describe this serious problem and the importance in university students of their attitudes and feelings about numerical self-efficacy and numerical relevancy on their numerical literacy competence. Surprisingly, these affective factors are stronger than many academic or demographic factors such as age, race, ethnicity, and gender. Their conclusions probably apply as well to secondary-level students and highlight an additional emphasis that is needed to address quantitative literacy.

### 3.2    Teaching the Culture of Science

Science educators have developed a greater awareness of how they, and the science community they represent, utilize a different language, different values, and even different social norms than the students they teach. The educator has to find ways to take advantage of the cultural resources brought to the educational setting to enable the linguistic, emotional, and cultural changes necessary to address what might appear to be simple unmotivated student behavior to learn science. Case studies (e.g. Kurth et al. 2002) provide valuable insights for how these cultural issues are embedded in the classroom. These cultural issues have strong influences on a student's acquisition of scientific discourse and on the interpersonal dynamics inherent in the context of students experiencing or "doing" science.

Thus, it should not be underestimated how difficult it might be for a student from various non--middle class backgrounds to interact, negotiate, collaborate, and to interact with peers as an in-tellectual partner as a part of the process of learning and doing science. Scientists functioning in educational roles should be aware of the potential challenges faced by these students in what might seem to be ordinary, culturally neutral, and even trivial interpersonal interactions in the science lab, for example. Barton and Yang (2000) discuss how socioeconomic and cultural issues apply to the science education of inner-city students and how the differences between home culture and school culture can adversely affect bright and talented students.

Research shows that adding the expressive arts to the combination of science, technology, engineering, and mathematics (STEM to STEAM), can be a powerful, culturally-centered engagement tool that can appeal to the unconventional learning styles of underrepresented and underserved students. The participation of minority STEM professionals also plays a powerful role in configuring the learning experiences of these students (Cummings et al. 2018). Unconventional approaches, such as combining art, music, play, and science in unusual settings can be most effective (Rosin et al. 2019). Building a diverse community in physics and astronomy is difficult and requires a long-term investment in new approaches (Stassun 2005; Shugart et al. 2018). STEAM programs may be one of them.

### 3.3 Formation of a Science Identity

These cultural adaptations play an important role in what is termed "agency" or empowerment, and in the concepts around how people form a science "identity." A science identity leads to a willingness to engage in ever more sophisticated interactions with science. An understanding of science identity formation leads to a much deeper understanding of how, why, and where people learn science, as well as the factors that can encourage or impede them (Bell et al. 2012,2014; Tzou et al. 2014). This social perspective is valuable in understanding how girls relate to science, particularly in the highly formative period when they are 10 to 14 years old (Tzou et al. 2014; Calabrese et al. 2013). The formation of this "Science Capital" is critical to a broad understanding of how educators can influence and change the science education ecosystem and is discussed later in more detail in section 7.

Songer and Linn (1991) describe how a student's view of science as either static (facts to be memorized) or dynamic (science ideas can develop or change) affects the way they learn. Unfortunately, too many students view science from the former perspective; their teacher's science instructional methods may even reinforce this belief. Alberts (2018) relates the perspective of Nobel Prize winning geneticist Lee Hartwell, who has argued for science education curriculum reform:

> *Nearly all science projects for schools involve repeating something that is well established so the goal is to get the right answer. That is not science. It is analogous to offering a history of art class to someone who wants a painting class.*

Scientists can play a key role in helping teachers and their students understand how science is done and to illustrate the excitement of scientific research and the value of our scientific culture, which is often misunderstood. These cultural differences also apply, on a different level, to how pre-service teachers (with their own subculture) react when exposed to the culture of science (Spector & Strong 2001). These cultural differences and impedance mismatches can apply as well when educational interactions occur between scientists and teachers, scientists and students, and among people of different nationalities. These possible misunderstandings have many implications for educational program design, and especially for programs brought to other countries for localization (Hofstede 1986; Hall-Wallace et al. 2002a, b).

### 3.4 Physics Education Research

Secondary school level astronomy education, particularly in the area of concept development, shares many basic traits with physics education, especially to the extent that one uses an approach that emphasizes abstraction and idealization. The research literature of physics education is extensive in this area (Beichner 2009) and addresses areas such as conceptual difficulties in fields such as optics (e.g. Goldberg & McDermott 1986) and the design of instructional materials that improve students' conceptual understanding and attitudes about science (Goldberg et al. 2010). Hestenes (1987) argues for a theory of instruction based on a mathematical model-centered instructional strategy that requires students to coordinate and integrate facts with scientific theory.

Redish (2014) has focused his research to provide a scientific basis to answers questions such as "What does it mean to understand something in physics?" and "What skills and competencies do we want our students to learn from our physics classes?" He also addresses the development of aspects of physics learning that represent a "hidden curriculum" and the culture of physics, by which an intuitive grasp of the subject is attained (Reddish 2010). These examples from physics education research remind us to reflect on the instructional strategies that we pursue in our field rather than focusing on the subject matter. When we perceive that our students are not grasping concepts, reflection on our instructional strategies is more valuable than blaming the students or merely repeating the lesson in a different voice.

## 3.5    Importance of Developing the Educational Skills of Scientists

For scientists to be comfortable in performing their different educational roles, they need training and support. Although educational roles for scientists may appear to be simple compared to their complex research roles, the scientist can be most effective if they receive some background training, practice, and reflection. The training program needs to focus on building an awareness of how scientists can engage with education stakeholders as a true partner, often playing a subsidiary or support role rather than leading the effort.

According to Stylinski et al. (2018), the research findings suggest that a comprehensive engagement training model, which incorporates learning theory, helps scientists build their own outreach strategies, provides opportunities to practice, and offers easy access to audiences, can have a sustained impact on disposition, perceived skills, and type of outreach conducted by scientists interested in deeper engagement with the public.

Training programs such as the NASA Astrophysics Ambassador Program (Cominsky et al. 2015) and the American Astronomical Society'sAstronomy Ambassador Program are examples of these training opportunities for astronomers interested in developing their skills and knowledge. The latter program plays an important role in developing the skills of astronomers who want to play a role in the educational ecosystem (Gurton et al. 2013; Fraknoi et al. 2014). These programs, if designed properly, can be highly effective at improving the educational attitudes,practices of scientists. They are also useful in helping scientists to develop program with greater levels of engagement.

There are many factors that characterize the forms of engagement by scientists including the depth, the motivations of all parties, the level of science focus, the type of participant, the different engagement actions, the type of interactions, and the outcomes desired. Storksdieck et al. (2016) has created an extensive typology for public engagement which includes these engagement categories and options. These differentiations can be useful to scientists in realizing the possibilities that exist for a wider range of interactions; they also indicate ways to insure higher quality, and more productive interactions with the public and education stakeholders.

## 3.6    Astronomy Education Research Literature

The field of astronomy education and astronomy education research has grown rapidly for the last several decades. In 2009, Lelliott and Rollnick reviewed astronomy education research on school students and educators over a 35-year period from 1974 until 2008. Wall (1973) reviewed doctoral or master's level studies in astronomy education, and Bishop (1977) assessed the state of U.S. astronomy education. Before 2001, there were no journals dedicated solely to astronomy education research.

The founding of the *Astronomy Education Review (AER)* by Fraknoi and Wolff in late 2001 (Fraknoi & Wolff 2001) was an important milestone that provided a central place for peer-reviewed publications, with 100 papers in the first five years (Fraknoi and Wolff 2007), The AER cove-red, for example, optics demonstrations (e.g. Birriel 2008), educational uses of online telescopes (e.g. Gould 2006), studies of assessment instruments (e.g. Sadler et al. 2010), reviews of research (e.g. Bailey & Slater 2003), and evaluation of astronomy podcasts (e.g. Gay et al. 2007). Of the more than 50 categories of papers (AAS 2013), some are distilled advice on better teaching practices (e.g. Fraknoi 2011), others are literature reviews (e.g. Bretones and Neto 2011) while others are evaluations of educational programs such as the micro-observatory program (e.g. Gould et al. 2006). Unfortunately, the Astronomy Education Review is no longer extant. However, research in this area continues at a vigorous pace. Longer studies, such as Ph.D. dissertations in astronomy education and related areas, have proven valuable as well (e.g. Fitzgerald 2015).

## 4.  WORKING WITH THE FORMAL EDUCATION SYSTEM

To be effective in working with the formal education system, the scientist must first sincerely appreciate that public school teaching is a demanding profession, with great responsibility, high stress levels, and many intellectual and emotional challenges. It requires a high level of competency for a teacher to function in a complex system that includes students, parents, administrators, curriculum designers and developers, and policy makers as they address ever-changing goals with limited time and monetary resources. The successful, expert teacher can only be a master if they can perform within this system. There is no solo performance as a career option for teachers. Teachers acquire their expertise from their specialized coursework, their supervised practice training, and from their years in different but demanding school environments. The long-term experiences of expert practicing teachers greatly supplant the pseudo-education expertise claimed by most everyone from being a school student who sat in a classroom for some 10,000 hours (Stigler & Miller 2018).

Teaching is a complex activity. Quality teaching is more than simply teaching using an assortment of the "best classroom practices." It involves a cycle of assessing the knowledge state of each student and class before and after instruction and creating learning goals. The teacher must be comfortable with a repertoire of strategies and approaches and know which ones to use and when. Then the teacher must effectively implement the most valuable approaches and strategies in creating learning opportunities for students. Good teaching also involves creating a quality relationship between the students and the teacher which encourages learning when learning is less fun and more difficult. Astronomers who want to "teach" might want to follow this same complex cycle if they are to be effective with students in classrooms.

Scientists play a variety of important roles in the formal education system, depending on their motivations, skills, experience, and temperament. Scientists at all career stages can help improve the conceptual knowledge and attitudes of both classroom teachers and students; aid teacher professional development; and help develop progressive, inquiry-based educational materials, including those that utilize problem-based learning. Scientists associated with space missions or other long-term projects have a unique opportunity to create and contribute to powerful, long-term educational efforts. For example, the Swift gamma-ray explorer mission supported fourteen years of education and outreach programs (Cominsky et al. 2014).

### 4.1  Components of Successful Scientist/Teacher Partnerships

Scientists who work in the classroom serve as novice public school teachers who can make use of their skills as expert scientists. Teachers are usually novice scientists, but experts in pedagogy. In a collaborative apprentice model, a partnership between scientists and teachers allows teachers to gain and share scientific knowledge, while scientists can gain valuable pedagogical skills. In this way, both can progress developmentally into higher levels of proficiency in their new skill sets (Ufnar et al. 2018).

The partnership will help teachers better understand the professional science activities of their astronomer partner, and the astronomer will gain a much greater understanding of the specific needs of schools and teachers (Bennett et al. 1998). A successful partnership allows multiple levels of engagement, depending on the time constraints and the motivation of the parties involved (Bybee & Morrow 1998), but should be mutually enjoyable and beneficial. The partnership develops over time as the partners open up to each other and learn to admit that there are gaps in their knowledge and understanding as they progress towards becoming an expert team with common goals (Salas et al. 2006, 2018).

### 4.1.1    Project ASTRO

Project ASTRO (Bennet et al. 1998), is one of the longest-lived sets of teacher-astronomer partnerships, having operated in the United States at over a dozen locations for over 20 years (Dierking & Richter 1995). It has been shown to be highly effective in urban schools and to positively influence the motivation and level of questioning from students (Rommel 2010, 2012). A key tool in the Project ASTRO collaboration is the use of two astronomy activity and resource books, *The Universe at Your Fingertips* and *More Universe at Your Fingertips*, which are used extensively by teachers and their astronomer partners as the basis of classroom activities (Fraknoi 1996; Fraknoi & Schatz 2000).

To assist the understanding of the different cultures of teaching and research, a Project ASTRO "how-to manual" was created and has been used extensively in Project ASTRO training (Richter & Fraknoi 1996). The activities have been translated into Spanish (Fraknoi & Schatz 2002) and are used in the southwestern United States as well as in Chile in these translated versions (Norman et al. 2003).

The experience in Project ASTRO, and other programs where scientists work in schools, is that scientists have a potent opportunity to help deepen the students' fundamental ideas and concepts of science, and even to answer questions like why astronomy is important to society (Rosenberg et al. 2013). If the astronomer develops a respect for the expertise of educators, communicates in a non-condescending, jargon-sensitive way, and collaborates rather than competes with educators (Morrow & Dusenberry 2004; Bybee & Morrow 1998), their effectiveness can be greatly increased.

### 4.1.2    Culture Differences between Scientist and Teacher

Each partner must be aware that scientists and teachers share tendencies towards behavioral attributes that may be quite different. Morrow and Dusenberry (2004) describe the cultural tendencies of scientists including an intellectual confidence or even arrogance, a competitive perspective, a critical perspective in their judgements, less social adeptness, a confrontational approach towards problems, and a desire to assign credit carefully for others' ideas. Teachers tend towards being less intellectually confident, being more collaborative and having an appreciative orientation, possessing good social skills and working around rather than confronting problems, and having a willingness to borrow good ideas freely from others without attribution. These different traits, though by no means universal, may create a gap between teachers and scientists that can make communication more difficult.

Despite these potential cultural differences, there are many advantages to be gained by both teachers and students when scientists partner with schools. Houseal et al. (2014) showed that student-teacher-scientist partnerships had positive effects on students' and teachers' content knowledge, attitudes toward science, and the pedagogical practices of teachers. Graduate student scientists who are program presenters in the classroom also can be quite influential if they introduce inquiry-oriented science materials or programs into the classroom (Hall-Wallace et al. 2002a).

### 4.1.3    Attributes of Effective Programs

In the most effective programs, teachers learn new science content and new ways to teach it,

and students gain new views of science and scientists. The graduate students also benefit from gaining teaching skills and a greater understanding of education and diversity issues, as well as gaining satisfaction and confidence from their experiences (Laursen et al. 2007). For primary grade education, a best-evidence synthesis shows that the most effective programs use inquiry-oriented approaches coupled with teacher professional development to help them use these approaches effectively (Slavin et al. 2012). For secondary school, a best-evidence synthesis supports programs that concentrate on teacher professional development, technology, and support for teaching (Cheung et al. 2017). Materials-focused innovations in education, such as textbook innovations, did not have as great an impact, especially when the adoption of books or kits by teachers did not include extensive professional development.

Scientists can assist teachers in making a transition towards these more effective approaches to science instruction and in using technology in teaching, another effective approach (Cheung et al. 2017). For secondary schools, which emphasize interdisciplinary problem solving and problem-based learning (e.g. Myers and Botti. 2000), scientists can be effective resources in both teacher professional development and in the design of instructional materials design, especially if they capture strong elements of inquiry and the process of science (Myers et al. 1997; Pompea & Walker 2017). The inquiry-based approach and perspective to science education is ubiquitous to science education and science education reform worldwide (Abd-El-Khalick et al. 2004).

Research on classroom visits by scientists indicate that having a female scientist can create a positive affect among students of both genders. More important, though, is the personal connection made by the scientist to the students, and the extent to which the scientist is perceived as "nice," "fun," and as someone who makes science interesting (Conner & Danielson 2016).

## 4.2 Research on Children's Ideas

Scientists are experts in the content knowledge of their field, the fundamental concepts and constructs that form the backbone of their specialty, an understanding of key experiments which have contributed to knowledge, and in the scientific process and enterprise. Students are not. Communicating that knowledge requires more than the astronomer remembering which approaches worked for them personally when they were young. An understanding of children's ideas in key science areas is essential.

An often-stated goal of astronomy education is to teach relevant science concepts, elucidate scientific reasoning, and to model the scientific process. Science educators have been grappling with the best ways to achieve these goals for many decades (e.g. Lederman 1992; Tobin et al. 1994; Lazarowitz & Tamir 1994; Amin et al. 2014), and their approaches will be valuable to any scientist wishing to work in a science classroom. The origin of the various science education international reform movements has its basis in finding better ways to achieve the above goals (Abd-El-Khalick et al. 2004). An understanding of how children perceive the world is key to any reform efforts in science education.

Driver et al. (2014) comprehensively reviews research on children's ideas about materials, the Earth in space, and physical processes such as electricity, magnetism, light, sound, energy, and gravity, as well as biological concepts and systems. This work is valuable for anyone teaching or developing instructional materials in any of these areas. Some notable research was done by Nussbaum (1976, 1979) and Sneider (Sneider & Pulos 1983; Agan & Sneider 2004) to address children's ideas on Earth's shape and gravity to guide teachers and developers of instructional materials. A similar effort was done for the phases of the moon and eclipses (Kavanagh et al. 2005).

A number of curriculum and teaching guides relied heavily on research on children's ideas as a basis for developing activity and inquiry units (Sneider et al. 2011). Examples informed by research are a teacher's guide on the seasons (Gould et al. 2000), another on teaching about the electromagnetic spectrum (Pompea et al. 2000), and instructional materials in optics (Sadler 2000; Pompea et al. 2007). These instructional materials or kits can form the basis of large, extended programs (Pompea et al. 2013).

A review of scientific thinking in young children (Gopnick 2012) highlights the importance of this research field on both education policy and the construction of instructional materials. It also reinforces the research foundation behind the rationale for using inquiry-based and innovative, interactive activities rather than more traditional approaches (Pompea & Blurton 1995; Myers et al. 1997).

## 4.3    Research on Learning Cycles

A starting point for science educators wishing to model better the scientific process was the development of learning cycle-based instructional approaches that emphasize scientific reasoning and the integration of conceptual understanding to allow students to construct knowledge. The approach of Karplus (1967) begins with an exploration phase where students explore a phenomenon and raise questions about observations or experiments. A concept introduction phase allows interaction among students, the teacher, and the instructional materials; in a concept application phase, students use the concept to solve new problems. This approach was extended in 1987 (Bybee 2014; Bybee et al. 2006) to the widely used 5E science instructional model with the categories of "Engage, Explore, Explain, Elaborate, and Evaluate" (Broggy et al. 2014, TEMI 2016). This model allows for multiple iterations that model the scientific process and allow for additional exploration as new evidence or data becomes available to either reinforce or challenge the initial conclusions. The TEMI project involved partners in 11 European countries and produced a highly useable set of classroom activities with worksheets that utilized this learning model (Loziak et al. 2015).

Adapted from Bybee et al. 2006, the model can be described as:

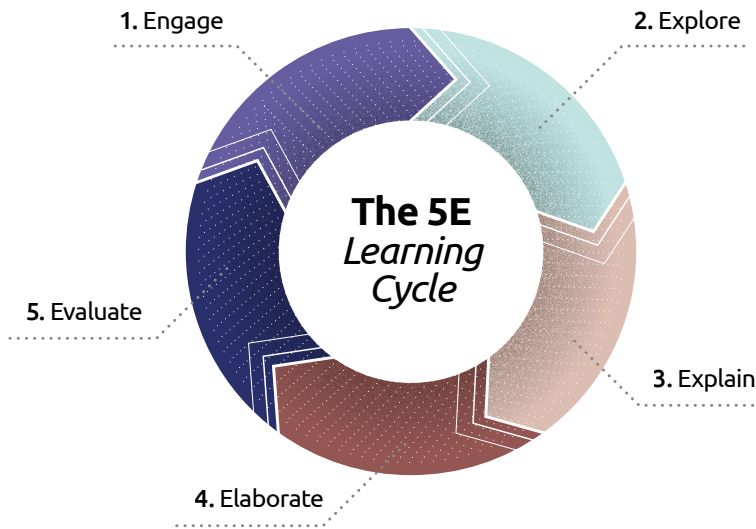

**Engagement:** The teacher, scientist, or a curriculum task accesses the learners' prior knowledge and helps them become engaged in a new concept through the use of short activities that promote curiosity and elicit prior knowledge. The activity should make connections between past and present learning experiences, expose prior conceptions, and organize students' thinking toward the learning outcomes of current activities.

**Exploration:** Exploration experiences provide students with a common base of activities within which current concepts (i.e., misconceptions), processes, and skills are identified, and conceptual change is facilitated. Learners may complete lab activities that help them use prior knowledge to generate new ideas, explore questions and possibilities, and design and conduct a preliminary investigation.

**Explanation:** The explanation phase focuses students' attention on a particular aspect of their engagement and exploration experiences and provides opportunities to demonstrate their conceptual understanding, process skills, or behaviors. This phase also provides opportunities for teachers to directly introduce a concept, process, or skill. Learners explain their understanding of the concept. An explanation from the teacher or the curriculum may guide them toward a deeper understanding, which is a critical part of this phase.

**Elaboration/Extend:** Teachers challenge and extend students' conceptual understanding and skills. Through new experiences, students develop a deeper and broader understanding, more information, and adequate skills. Students apply their understanding of the concept by conducting additional activities.

**Evaluation:** The evaluation phase encourages students to assess their understanding and abilities and provides opportunities for teachers to evaluate student progress toward achieving the educational objectives.

The astronomer who works in classrooms or who wishes to develop laboratory or other classroom materials should be familiar with this 5E model.

## 4.4    Research on Preconceptions, Naïve Theories, and Misconceptions

Extending our understanding of children's ideas of how the world works is an understanding of the role of student preconceptions, naïve theories, and misconceptions on a student's ability to construct new knowledge (Wandersee et al. 1994; National Research Council 1997). Extensive research has been done in this area; an excellent bibliography is available from Carmichael et al. (1990). Although the prior conceptions of students are often viewed as impediments to learning new concepts, they can also serve as resources for cognitive growth (Smith et al. 1994). The awareness that bright students (all students!) in a class might have preconceptions or wrong ideas based on enduring misconceptions has promoted a reexamination of the effectiveness of traditional teaching methods. The film "Private Universe" examines a number of misconceptions such as the cause of the seasons and their persistence even in the face of focused instruction (Crouse et al. 1989). The number of categorized astronomy-related misconceptions has grown to more than 1700 through careful study (Comins 2001).

The persistence of misconceptions has been an area of significant concern since unlearning old ideas may be more difficult than learning new concepts. Teacher competency may depend on their knowledge of student errors (Shulman 1986; Sadler et al. 2106) and their ability to uncover and diagnose student preconceptions early in the learning process (Treagust 1988). Similar efforts to understand what students already know have been undertaken at the university level in an attempt to inform teaching through a diagnosis of student knowledge at the beginning of a course (Hufnagel 2001; Deming 2002; Libarkin et al. 2011).

Research on how best to address preconceptions, misconceptions, or naïve theories in science education is important (e.g. Stavy & Tirosh 2000). One approach is to use cognitive dissonance to reinforce the need for conceptual change (Stepans 1996), while another approach focuses on formative assessment probes to help the instructor understand better the precise ways that students think about key ideas (Keeley & Sneider 2012). In Stepan's conceptual change model, students are encouraged to confront their own preconceptions and those of their classmates. As they become aware of their own preconceptions about a concept, they make predictions and commit to an outcome, such as an experimental outcome. Students then work to resolve the conflicts between their ideas and their observations in order to accommodate and extend the new concept.

This process extends the conceptual change model through an extension of the learning cycle model of Atkins and Karplus (1962). Key to this effort is an appreciation of how knowledge and meaning are constructed by students through active engagement in meaning-making (Jones & Brader-Araje 2002) using language to create the individual's conceptual ecology and to mediate higher order thinking, serving as a tool in the meaning-making process (Vygotsky et al. 1994). An evaluation of misconception research from a constructivist perspective argues that student conceptions and misconceptions play productive roles in the acquisition of expert knowledge (Smith et al. 1994). Culture also plays a strong role in the creation of mental models and should not be ignored (Nobes et al. 2003; Siegal et al. 2011).

## 4.5    Scientists and the Professional Development of Educators

Astronomers can play a key role in the training and professional development of science educators (sometimes called "teacher training"). This can be helpful to teachers at any career phase, from pre-service development work while the future teachers are at the university (Hemenway 2005) to broad enhancements in the knowledge and techniques of practicing educators (National Research Council 1996). Professional development is equally important to science educators working in informal education settings such as museums, after-school programs, science centers, and planetariums. The comfort level of primary school teachers towards the subject plays a key role in the educator's decision as to how much time they will devote to teaching astronomy or whether they will teach it at all (Chastenay 2018)!

It might be presumed that the main role of scientists in the professional development of educators is to increase the educator's subject matter knowledge, presuming that the teacher understands the teaching pedagogy related to their subject area. This teaching role is valuable but rather limited, as teaching skill is not related to knowledge level. If this were true, college instructors with doctorate degrees in their subjects would be the best teachers of all. In practice both the content knowledge and additional practical knowledge on teaching that content (pedagogical content knowledge) need to be improved.

### 4.5.1    Pedagogical Content Knowledge

The key question is this: What specialized knowledge is essential for quality teaching? Shulman (1986) framed the discussion around the concept of knowledge that is developed by teachers that helps others learn. This form of knowledge is subject specific and combines content knowledge, pedagogical knowledge, and contextual or situational knowledge in a combination that generates quality teaching. Pedagogical Content Knowledge (PCK) combines these three forms. Helping teachers improve their PCK should be a goal for the scientist.

Improving the educator's PCK can be done by the scientist using her or his experience to offer insights into student misconceptions, the field's historical context, and the determination of exemplary instructional materials (van Driel et al. 1998, 2014). This combination can greatly enlarge the educator's perspective and teaching effectiveness. Research on PCK is ongoing (e.g. Carlson & Daehler 2019) and is highly relevant to improving the training of secondary school teachers (Lederman & Gess-Newsome 1999). Various programs have been explicitly designed to improve the PCK of astronomy instructors (Prather & Brissenden 2008).

### 4.5.2    Basics of Educator Professional Development

Effective professional development must address the learning goals of the audience and, as such, relies on several key principles for efficacy (Loucks-Horsley et al. 2010). These principles from Loucks-Horsley et al. include creating the opportunity for teachers to examine their practices, to enhance their content and pedagogical content knowledge, and to have time for collaboration with colleagues and experts. Ideally, the program must also provide transformative learning experiences that improve instructor competency and create new ways to improve student learning. This transformation often leads to cognitive dissonance, and sufficient time and a supportive environment must be provided for the teachers to work through the dissonance. Providing support for improving pedagogical content knowledge is particularly important since this knowledge includes the general orientation towards science teaching and the science curriculum, an understanding of student knowledge and beliefs, and a knowledge of assessment and instructional strategies (Gess-Newsome & Lederman 1999; van Driel & Berry 2010; Pompea & Walker 2017).

The length of the professional development experience is also critical. While most educators have participated in short-term workshops, research indicates that more extended professional development programs (50 or more hours) are needed before teachers substantially change their practice (Wei et al. 2010). Videoconferencing can be effective in bringing professional development workshops to teachers directly in their communities rather than requiring them to travel to a remote location (Hemenway et al. 2009).

### 4.5.3    Challenges of Contributing to Professional Development

One of the most rewarding and challenging roles of a scientist is to provide professional development for educators. On the surface, nothing could be easier: teachers of all grade levels need to improve their content knowledge and understanding of how science works. Scientists are experts at both. With sensible communication skills, success can come quickly and easily. This novice perspective should be contrasted with how an expert would approach professional development. First they would take on the challenge of finding the right teachers for the subject matter of interest. Teachers at each grade level have standards (national, regional, local) to adhere to and curricula to follow, often with high-stakes testing (e.g. the teacher might lose their job) to establish how successful the teacher has performed their job.

The workshop leader needs to find the right grade level for the topic and to understand and assess the current general knowledge of the participants. Then a location and time must be found (usually on a weekend or holiday) and a way to communicate with potential attendees must be established. The teachers expect to receive some credit (which may count for their recertification) for attending or perhaps even to get a monetary stipend. When delivering the content, the leader must understand how the content will be integrated with the pedagogy and in particular with pedagogical content knowledge.

The workshop itself is expected to reflect best practices such as hands-on activities and guided inquiry. The activities used must be well-tested, age appropriate, and be responsive to the standards, if they are to be used after the workshop. There must be time for teachers to try out what they might teach and for them to discuss concerns or issues with implementation. If the activities require special supplies, these need to be furnished for the workshop and perhaps a mechanism provided to loan these supplies to the teachers. After the workshop, the teachers will need support to implement teaching any new topics or in teaching old topics in new ways. If the topic is not normally taught until later in the school year then it is critical that support be provided at that time. The workshop should be evaluated to find ways of doing it better. The long-term goals of the workshop should have been established at the beginning. An evaluation of whether those goals were met should be done. Since the goals are probably longer-term goals, the evaluation period will need to be an extended one.

All of these issues are important to address, and many of them can be more easily handled by partnering with the school district, by using master teachers, and using an experienced evaluator to guide the process of setting up and implementing a successful professional development workshop. Although not all of these requirements apply to every country, our experience in many countries indicates that most of them do.

In summary, the research-based characteristics of professional development for teachers has these attributes (adapted from Cormas & Barufaldi 2011):

— The teachers' discipline-specific knowledge is increased.

— Teachers understand how students learn and the effective teaching strategies that can be used within a specific discipline.

— Teacher effectiveness and student achievement outcomes are used to determine whether professional development has worked.

— Professional development is on-going and requires resources (money and time).

— The program uses effective teaching strategies such as inquiry.

— Teachers provide input into the professional development design.

— The professional development is engaging and relevant.

— The program involves collaboration and generates further collaboration on projects.

— The professional development program treats teachers as professionals and promotes teacher self-reflection.

— Overall, the program increases the teacher's ability to meet the needs of diverse learners.

Scientists can help a professional development program fulfill many of the characteristics described in this list. The training and background of scientists can make them a valuable member of a science education professional development team.

## 4.6    Development of Instructional Materials

With their deep subject matter and contextual knowledge, scientists can play a key role in one or more stages in the development of instructional materials. Indeed, scientists have played important roles in many major curriculum reform projects in physics and astronomy such as the PSSC Physics and Project STAR (Rudolph 2002) and in the authoring of primary and secondary school textbooks (Pasachoff 2005). The contributions of scientists can include the development of activities, demonstrations, laboratory or telescopic investigations, paper and pencil labs, museum displays, exhibits, and new student telescopes. Astronomers have compiled practical guides on effective teaching techniques (Pompea 2000; Pompea et al. 2002) and have written scientifically accurate books for children (e.g. Bennett 2014, 2006). Astronomy activity books of high quality have also been produced (e.g. Schatz 1991, Schatz & Fraknoi 2016).

The developmental stages of these longer-term projects in which scientists can play a functional role include problem identification, needs assessments, idea generation, brainstorming and strategizing, creating partnerships, writing funding proposals, reviewing current or new materials, writing, testing new approaches or materials, and assessing educational program effectiveness. A key role well-suited to scientists is in evaluating educational materials. Russo et al. (2015) proposed the use of a peer-review workflow to improve the standards of quality, visibility, accessibility, and credibility of astronomy education activities. There is little need for another web repository of educational materials but there is a large need for a mechanism for peer-reviewing and publishing high-quality astronomy education activities in an open-access format.

The production of high-quality educational materials has been made possible by astronomers who played key roles. For example, the IAU *astroEDU*, a peer-review platform for astronomy education activities and the *Teach Astronomy* website (http://www.teachastronomy.com) have many high-quality outreach resources created by professional astronomers. There are also materials suitable for general outreach that have a broad coverage of astronomy news and are suitable for children interested in astronomy. Some examples include *Space Scoop* (Tran & Russo 2018) or *Science Daily*. There are also extensive, high-quality websites and online textbooks such as the one created by Impey (2013) and by Fraknoi, Morrison, and Wolff (2019).

Educational materials on eclipses were developed, featured in teacher resource catalogs, and were widely used for the 2017 total solar eclipse (Schatz & Fraknoi 2016). Scientists have played roles as project directors for many national astronomy and technology education development projects (e.g. Pompea et al. 2005, 2006, 2013) and in both the International Year of Astronomy 2009 and the International Year of Light 2015. They have also played a key role in the IAU100 educational projects, centered on the 100[th] anniversary of the IAU. Scientists can be particularly helpful when developing problem-based learning materials (Schwartz et al. 2001) which have a close correspondence to authentic scientific problem-solving processes (Hong, 1998; Hong & McGee 2000; Shin & McGee 2002; Myers et al. 1997; Pompea & Walker 2017).

Educational materials and programs have been developed to serve visually impaired visitors using braille guides, embossing, and the use of tactile models (Bartus et al. 2007). For example, a tactile model of the Sun was created in a collaboration that involved an astronomer, an educator, and a sculptor in order to make astronomy more accessible to visually impaired students (Isidro & Pantoja 2014). Other astronomy education resources for visually impaired visitors and students have been developed by Grice (2006) including ones addressing multiwavelength astronomy using tactile books (Grice et al. 2007). Science data can be converted to sound, providing an additional manner in which auditory senses can communicate patterns that may occur (Keller et al. 2003; Quinn 2001).

The development of instructional materials that are responsive to teacher needs and that can be integrated into the current curriculum or used as supplemental materials can be challenging. Wiggins and McTighe (2005) describe a process of "backwards design" that has proven to be highly successful, when planned and executed properly, in the design of curricula, assessment, and instruction. Many educational projects and development teams use this approach.

## 4.7 Educational Use of Astronomical Data

Programs that use astronomical data include citizen science projects (described in another section in more detail), research experiences for high school or undergraduate students, programs that involve using telescope or archival data, programs that emphasize laboratory research, as well as those that involve computational research and simulations. These educational programs can use simulated science experiences (e.g. Pompea & Blurton 1998) or science-related games (e.g. Li & Tsi 2013) to facilitate learning. Often it is the role of the scientist helping to design these educational programs, simulations, or games to make sure that the science concepts and process are carefully and accurately represented.

### 4.7.1 A Cognitive Apprentice Perspective

Many astronomers are drawn to creating or helping astronomy education programs which have a basis in the educational research fields of cognitive apprenticeship and authentic science practices. Cognitive apprenticeship may be viewed as a modern version of teaching a trade that allows the expert to model a skill and explain what they are doing and thinking while the apprentice notices important behaviors and develops a conceptual model of the processes involved (Collins et al. 1989; Brown et al. 1989). Authentic science practices can increase the motivation of students to learn, and also to frame that learning, in the epistemology or knowledge structure of science. Exposure to science data in programs like teen science cafés can be effective in motivating students to learn more about how modern "big data" astronomy research works (Walker & Pompea 2018).

Fortunately, there is a rich body of research literature on how to effectively implement the notions of cognitive apprenticeship and authentic science practices. Collins (2006) reviews the research on cognitive apprenticeships; Edelson and Reiser (2006) discuss in detail the design challenges and strategies for making authentic science practices accessible to learners. Citizen science projects have a wide reach throughout the scientific fields and are evolving rapidly (Bonney et al. 2009, 2014). Volunteers in citizen science projects can collect expert-level quality data if the appropriate steps are taken to ensure the proper training of the participants, the creation of appropriate protocols, and careful monitoring of the output (Danielsen et al. 2014). In astronomy citizen science projects, close collaboration between the professional and amateur community can be most effective (Walker et al. 2011). This is particularly true if the projects involve work in unpopulated scientific niches. Marshall et al. (2015) described how citizen scientists can play strong research roles both as "observers" and "classifiers."

### 4.7.2 Collaborative Research Teams

Collaboration programs that team researchers with high school students and teachers have been very successful but have inherent challenges. A comprehensive review is given by Fitzgerald (2015). Early programs of this type were sponsored by organizations in the U.S. such as the Research Corporation (Tucson, Arizona) which created a long-term program for partnering high school teachers nationwide with university researchers beginning in about 1988 (McCarthy & Lockwood 2013). The success of this program and its impact on teachers and students led to many programs funded through the National Science Foundation (U.S.) such as the NOAO *Re-* *search Based Science Education* program for students and teachers (Jacoby et al. 1998). The main requirement for these programs to succeed is that all participating parties must benefit. The scientist must see some advantage in having students work on his or her project. The teachers should realize that they, and their students, will develop new skills and content knowledge. All parties must feel that their investment in the project has personal rewards and gains (Ledley et al. 2003).

### 4.7.3    The Role of Remote and Research Telescopes

Examples of these programs include the MicroObservatory program (Dussault et al. 2018) and the Las Cumbres Observatory program (Brown et al. 2013). The MicroObservatory was one of the first networks of automated remote telescopes that were dedicated to educational use. From this program, key attributes of value included the importance of students having control of the telescope and ownership of the images, the allowance of students to fail "constructively," easy and equitable access to the telescopes, and a high level of authenticity present in the use of the telescopes (Sadler et al. 2001) After ten years of use of the MicroObservatory, and after an examination of the results of 475 inquiry-driven projects in all 50 states in the U.S., the project reported significant gains in the understanding of core concepts in astronomy and physical science, use of inquiry skills, integration of mathematics into models, and in the basic conceptual understanding of how telescopes function (Gould et al. 2006).

Other programs have used research-grade telescopes to encourage collaboration among professional astronomers and teachers. These include the Hands-On Universe program (Pennypacker & Barclay 2003; Pennypacker & Asbell-Clarke 1996), the Teacher Leaders in Research Based Astronomy program (Pompea et al. 2005), the Spitzer Space Telescope Research Program for Teachers and Students (Spuck et al. 2010), and the NASA IPAC Teacher Archive program (Rebull et al. 2018), to name a few. Fitzgerald et al. (2014) provides a comprehensive review of 22 high school astronomy student research programs that have operated over the last two decades. Although nearly all of these programs encourage genuine inquiry, observations, hypothesis testing, and other aspects of the astronomy research process, it is useful to apply a deeper analysis of what constitutes an authentic, inquiry-driven research experience and to reflect on the framework for epistemologically authentic inquiry in school-based research programs (Chinn & Malhotra 2002).

### 4.7.4    Tools to Access Archival Data

Other resources like *Google Sky*, *Stellarium*, and *The Worldwide Telescope (WWT)* have altered how astronomical data may be used and viewed. The new, large, and available astronomical data sets are revolutionizing how data is used in astronomy education (Borne et al. 2009). For example, the WWT is a free, powerful, interactive, visual browser of multi-wavelength data that can be used with scripts. Because of its flexibility as a display platform, it can be used equally well by students and researchers and is widely used as an educational tool (Goodman et al. 2012). There are many efforts underway to create better tools for astronomy research large amounts of archival data.

## 5.    INFORMAL ASTRONOMY EDUCATION PROGRAMS

The Association of Science Technology Centers (ASTC) estimated that, in 2017, there were over 120 million visits to its science center and museum members around the world, and 70 million visits to the 386 science centers and museum members of ASTC in the United States (ASTC 2017). In-person yearly visits to museums in the United States surpasses the attendance of all major league sporting events and theme parks combined (AAM 2007). These "informal" learning or "free choice" education programs and settings play fundamental roles in the science education ecosystem and in providing venues for learning science and encouraging an interest in science and science careers (Falk 2001). A useful high-level summary of how science learning takes place in informal environments and which also includes vignettes of successful projects was produced by the U.S. National Research Council (NRC 2009).

The community impact of science centers and especially the large urban science centers that serve diverse audiences is quite substantial (Falk & Needham 2011), indicating that partnerships with science centers can be highly productive in reaching heterogeneous audiences. It is very common for astronomers to conduct astronomy education work in these afterschool and out-of-school settings.

## 5.1    Informal Settings

Astronomers often work in or with observatory visitor centers, museums, libraries, planetariums, and science centers. They participate in educational activities with community groups, professional societies, electronic media producers, and at national parks, science festivals, and street fairs. Collaborations with public libraries have been shown to be particularly effective (Smith et al. 2012). A system-wide perspective of the larger science education community can give many useful insights into the partners and structure of the network as well as the strengths and weaknesses of the system. These insights can help researchers, practitioners, and policy makers improve the overall quality of science education delivered by the system (Falk et al. 2015).

Astronomers can conduct programs at civic clubs, restaurants and pubs, participate in star parties, and contribute at large astronomical events such as those associated with solar and lunar eclipses, meteor showers, viewing of bright comets, or transits of Mercury and Venus. The programs they participate in can be topical and related to a space mission (e.g. landing on an asteroid), be specific to the release of an astronomical image (e.g. Hubble Deep Field), or can be timed to the announcement of an astronomical discovery (e.g., the LIGO detection of gravity waves from merging black holes or the first image of a black hole).

## 5.2    Challenges to Participation

Out-of-school programs often emphasize inquiry-based learning, which may be lacking in poorer school districts with minimal equipment or resources for use in STEM classes. Studies of practicing scientists have shown that these inquiry-learning experiences at home and in their communities have played sizable roles in encouraging their science interest and understanding (Christensen et al. 2015). Out-of-school activities such as observing stars, tinkering with mechanical or electrical devices, or reading and watching science-related materials play an important role in retaining astronomy interest throughout high school. Astronomy activities such as looking at stars plays a stronger role in females than in males in contributing to the retention of an interest in an astronomy career during high school (Bergstrom et al. 2006).

These settings are often more accessible to both the astronomer and to audiences which include underrepresented groups. However, much work remains to be done in this area to attract diverse audiences to museums and science centers (e.g. Dawson 2014). Ethnicity, geography, and socioeconomic factors are important in understanding how youths and others vary in interest and exposure to informal science activities. For example, rural youths report rates of lower participation than urban youths, and Latino/a youths generally have less out-of-school science experiences than non-Latino/a youths, especially in urban areas. The level of experience with out-of-school science programs or activities correlates with the available resources (e.g. number of books) at home (Hill et al. 2018). Immigrant youth are marginalized in most societies but can benefit from out of school programs. The designer of such programs should develop a deeper understanding of the distinct participation patterns of immigrant youth in their community who come from different backgrounds (Peguero 2011).

## 5.3     Best Practices: Are they Unique to Informal Settings?

This section will summarize the research on best practices for astronomers who are involved in informal education (defined as free-choice learning or participation in educational settings and programs that do not have the compulsory nature of classroom education). Generally, the same evidence-based fundamentals and best practices, as already described, apply just as strongly to informal settings. In the past, informal education was viewed as being distinct and different from formal education. Formal education was characterized as adults (teachers) leading programs that were highly sequential and coherent, and that emphasized concept development and vocabulary retention. Informal education was often viewed as the opposite, with an emphasis on fun, student-centered activities that did not emphasize learning sequences.

The current understanding is that there is a false dichotomy between the design differences of informal and formal programs. Both need to highlight a student-centered approach where adults play an important role. Informal programs are not stand-alone items but are part of the overall STEM skills development journey for youths. Both formal and informal programs serve to encourage them, expand their perspectives, and hopefully create a deeper level of engagement with STEM learning. The low-stakes, test-free environment of an informal setting is conducive to experimentation, exploration, and risk-taking, all of which are critical features of STEM learning (Volmert et al. 2013).

## 5.4     Science Centers and Museums

There is a rich body of free-choice learning research literature that covers why, how, what, when, and with whom people learn in museums and other informal environments. This research also covers the many dimensions of the museum experience and the current reinvention of the museum to support lifelong, life-wide, and life-deep learning (Falk & Dierking 2018). The literature also addresses the evaluation of public understanding of research projects and research initiatives (Storksdieck & Falk 2004), which can play a critical role in the funding and establishment of new projects, space missions, or even observatories.

One approach to involving scientists with museums is at the core of the Portal to the Public project (Selvakumar & Storksdieck 2013). This project provides a model for bringing science to informal settings by bringing community scientists and public audiences together. It represents a flexible and scalable approach that is likely applicable to every field of science. It includes a well-designed professional development experience for scientists that helps them develop communication skills and strategies that support inquiry and greater personal connection with the public. It also encourages collaboration with museum staff to gain an appreciation of each other's culture—a key feature in the design of high-quality learning experiences. The framework of a collaborative partnership between the research center and educational institutions is a powerful one (Pompea & Hawkins 2002). The partnership can effectively expand the broader impacts of the research into much more extensive citizen engagement activities (Alpert 2009).

Students visit museums to have novel experiences that can assist their learning. The research literature on student field trips applies as well to scientist-assisted visits to research sites by school groups. The value of these visits to unique settings is their ability to promote discovery and exploration, and their ability to create original experiences for the student. While field trips can be inspiring, they are less effective as a setting for concept development (DeWitt & Storksdieck 2008).

## 5.5     Planetariums

The opening of the first modern planetarium system by Zeiss in Munich (Germany) in 1925 was the beginning of the planetarium's current and important worldwide role in astronomy education. There are more than 3,200 planetariums of all sizes worldwide, including fixed and portable domes. The International Planetarium Society (IPS) has nearly 700 members from 35 countries. There are more than 20 regional and national planetarium associations from all over the globe that are affiliated with the IPS. Planetariums are effective astronomy teaching tools (Brazell &

Espinoza 2009) and can help children learn about patterns, representations, and lunar phenomena (Plummer & Small 2018). They have proven valuable as part of an integrated learning experience on celestial motions that combines school and museum aspects (Schmoll 2013).

Astronomers often teach in planetariums or collaborate with them in public events. In the future, the tremendous visualization capabilities of planetariums may make them ideal research venues for scientists and citizen scientists in examining and understanding large astronomical data sets (Wyatt 2007; Subbarao et al. 2013). Planetariums can also be used to stream public science lectures to other planetariums. Many planetariums have software that can use enormous astronomical catalogs and data, creating opportunities for celestial voyages through the universe and the three-dimensional exploration of astronomical objects like the Orion Nebula. Examining the many properties of the European Space Agency's Gaia dataset in a planetarium setting would be one example of how planetariums can bring enormous amounts of data to the dome where it can be examined by researchers and the public alike. Efforts to streamline the process to bring "Data to Dome" have been successful in reducing the effort that it takes to bring scientific data to actual visualization in the planetarium (Sabbarao et al. 2013).

Small planetariums and portable planetariums with more limited visual projection capabilities (and thus less impressive star fields, visuals, or movies) can still be highly effective as educational tools since it is the program, not the special effects, that often determines educational efficacy. Planetarium educators have high quality training guides available to them (Friedman et al. 1980) that originated in small planetariums such as the Holt planetarium at the Lawrence Hall of Science in Berkeley. Interactive programs, such as those in the Planetarium Activities for Successful Shows (PASS) series, have proven highly effective and are easily available to educators worldwide (Gould 2019). Planetariums are also useful in bridging the culture gap between scientists and the local community, with efforts in Hawai'i at 'Imiloa Astronomy Center in Hilo being most notable (Ciotti 2010).

## 5.6    Observatory Visitor Centers

Observatories and other astronomy research facilities, such as LIGO often have visitor centers that host a variety of exhibits and programs that can appeal to different audiences. These sites also host teacher professional development workshops, often using the exhibits for demonstrations and exploration. The major existing observatories of the world and those new ones being built (Krisciunas 1988; Wolff 2016) provide a wide range of experiences in the forms of exhibits, guided tours of the telescopes and research areas, solar viewing, twilight programs, bookshops and stores with astronomical items, and star parties with telescope viewing. Some have opportunities to talk with scientists. The programs are designed for visitors of all ages, including those with disabilities. For example, at the Kitt Peak National Observatory Visitor Center (USA), there are educational exhibits and retail operations, tours of telescopes through a vigorous docent/volunteer program, educational programs for visiting school classrooms and the general public, and a popular solar and nighttime observing experience program for both the general public and advanced amateur astronomers (Isbell & Fedele 2003). These visitor center programs are designed to serve the stakeholders and certainly have an important role in public relations as well as in education (Finley 2002).

Daytime observatory visits to see astronomical equipment or to make simple observations can be powerful motivators for primary grade students (Columbo et al. 2010). Unusual modes of transport, such as horses and paragliders, have even been used to bring telescopes to isolated locations (Seidel et al. 2016). These efforts have been valuable in bringing science directly to primary grade schools and in providing hands-on immersive activities not normally available, as well as the opportunity to interact with scientists (Roden et al. 2018). The interactive traveling science center, telescope, or experiment station is part of a long-standing "mobile museum" tradition (Rees 2016) that has proven its effectiveness in many venues.

The field of "visitor studies" is a broad and important area for every visitor center to examine, with a good overview by Hooper-Greenhill (2006). The field has a professional society (Visitors Studies Association) and journals (e.g. Visitor Studies) that can provide a wealth of relevant information to the educational experience designer. Simple, measurable items such as how visitors spend their time can be quite important to a general understanding of visitor learning and how visitors na-vigate through your exhibitions (Serrell 1997). There are many excellent resources available for understanding exhibition design (e.g. McLean 1993) and for a broad understanding of the experiences of visitors (Falk 2016). These resources will be especially valuable for visitor centers that are creating new programs, expanding their facilities, or trying to attract new audiences.

## 5.7    Astronomy Camps

Observatories also play an important role in hosting astronomy camps. The audiences for these camps are generally secondary school students, but they can also include university students, adults, leaders of youth programs, or teachers. These camps – such as the International Astronomical Youth Camp, are immersive in nature and create powerful experiences for their attendees. If properly designed, they can create opportunities for authentic research as well. Astronomy campers can learn how to formulate science questions for investigations, train on using high quality telescopes, and utilize powerful data tools to have a full range of research-related experiences.

The University of Arizona Astronomy Camp (Fields 2009) is one notable example (Fields 2009) of how to create an immersive, highly educational, authentic research experience, such as described by Chinn and Malhotra (2002). Directed by an astronomer, high school age students live for eight days in the astronomers' dorms and conduct research using professional instruments on four telescopes. The camp is staffed by undergraduate, graduate, and postdoctoral astronomy students. In this well-designed, highly intentional program, students develop strong and enduring friendships, an enhanced sense of personal autonomy, and deep science knowledge. The camp is a safe "affinity space" where students can develop under the supervision of caring and knowledgeable staff. As such, this camp is successful in nurturing the formation of a science identity in its participants, and many graduates have taken advanced science degrees and are now professional astronomers.

## 5.8    Amateur Astronomers and Astronomy Clubs

Astronomy education is greatly benefitted through the leveraging efforts of networks, partnerships, and intermediaries such as amateur astronomy clubs (Manning et al. 2008). It is interesting to note that amateur astronomers do not uniformly participate in educational roles. Some are quite active while others shun this role. Astronomers working with amateur astronomers should understand the factors contributing to amateur astronomers participating in education, which include both individual and club-related variables (Yocco et al. 2012). Amateurs in clubs do nearly double the amount of outreach of unaffiliated amateurs and are more knowledgeable about astronomy as well (Berendsen 2005). Supporting the growth of a culture of public outreach in these amateur interest and affinity clubs is critical (Berendsen et al. 2010) as these clubs are widely established. To increase the participation of amateur astronomers in outreach, work must be done to remove the barriers to amateurs joining clubs. It is also helpful to encourage more formal or informal training in astronomy to help amateurs feel more comfortable in educational roles. Networks of astronomy clubs such as the Night Sky Network and Astronomers without Borders have played important roles in programs such as the International Year of Astronomy 2009 (Berendsen et al. 2008).

Amateur astronomers are also key partners in teacher-scientist partnerships like Project ASTRO (Fraknoi et al. 1996; Fraknoi & Lalor 2000). Perhaps, surprisingly, amateur astronomers are perceived by classroom teachers to be as effective as professional astronomers (Gibbs & Berendsen 2006). A how-to guide by Richter & Fraknoi (1996) describes quite well how to form successful partnerships between amateur or professional astronomers and teachers.

## 5.9    Star Parties

Star parties can happen in almost any venue. Sidewalk astronomers set up small telescopes in places where people congregate. On urban streets, or outside shopping malls or stadiums, a telescope and its operator can attract multitudes for an exciting view of the Moon or Jupiter. A large or fancy telescope is not necessary; small, portable telescopes, such as Galileoscopes, have proven effective in urban environments and at rural star parties (Pompea et al. 2010, 2013). Some star parties are centered around large public or government events (Lubowich 2010; Pompea & Norman 2009) and get media attention as well. Research by Wenger (2011) shows that star parties are effective venues for engaging the public about astronomy and that an understanding of the dynamics of star parties can be most useful.

A star party is an opportunity for public engagement and longer discussions. Questions at star parties often are about astrology and UFOs. There are many resources available to familiarize astronomers on scientific responses to these topics (Culver & Ianna 1988; Fraknoi 2003, 2008), as well as other topics such as the effect of the full moon on behavior (Kelly et al. 1985). However, it is important to remember that while a factual response or logical argument might be appropriate in order to answer questions of this nature, it is also critical that scientists be sensitive to the "culture wars" that are present in any discussion or museum exhibit (Conn 2006) where pre-existing beliefs are strong.

## 5.10    Science Cafés

The Café Scientifique model for teens has proven to be a highly effective way to connect with secondary school students who may have only low to moderate literacy in science. It usually consists of a combination of a teen audience, food, a brief talk by a scientist, discussion, and an activity. By illustrating the value and fruits of science research and its connections to their lives, science cafés have proven effective in framing and illustrating how science works and giving a view of a scientific life. How the science presenters communicate with teens is critical; interactivity is vital. Extensive preparation is necessary, including several dry runs of the talk (Mayhew & Hall 2012; Hall et al. 2013). The café is often organized around the "most important thing," an essential and provocative ideas that can be reached through a story or strong narrative.

The teen café model recognizes the ineffectiveness of the information deficit model where experts with information feel the need to communicate that information to teens who lack it (Nisbet & Mooney, 2007; Nisbet 2009). Teen cafés can have specialized functions, such as those that emphasize modern computational astronomy tools and projects that use massive amounts of data (Pompea & Walker 2018). These cafés may prove useful in attracting teens to careers in computational astronomy or to the use of the large data sets generated by telescopes such as the Large Synoptic Survey Telescope (LSST) (Walker & Pompea 2018; Olsen et al. 2018).

## 5.11    Science Festivals

Science festivals and street fairs are useful venues for astronomy discussions or safe solar observing, often using solar projection telescopes such as the Sunspotter telescope developed through Project STAR at Harvard. Contests and competitions are also successful in generating a greater public interest in science. These can be based on making things (perhaps using maker spaces), creative problem solving, Astronomy Olympiads where students answer questions, or naming contests (Montmerle et al. 2015), such as the 2019 IAU Name ExoWorlds contest.

Science talks and discussions in restaurants or pubs have proven successful as well as informal educational opportunities; they require little funding to implement them (Rice & Levine 2016). The main requirement is that there be adequate time for meaningful interactions between the presenter(s) and the audience, and that the talks are engaging and compelling while staying between ten to twenty minutes in length. Sometimes small prizes (such as science books and posters) are used to encourage interaction. These events can be used to communicate current science research, encourage a sense of the humanity present in science, illustrate the diversity of scientists and their personalities, and encourage networking. They are also great practice venues for scientists who want to improve their communication skills with the general public.

## 5.12    Preschool and Early Primary Grade Programs

Research astronomers working with early primary grade or even preschool children (younger than age 5) may seem like a complete impedance mismatch, but it is not. It is an area of great importance and significant new research. Why is it important? First, many scientists have recognized the importance and value of working with younger children, such as their own children and grandchildren, and those of their friends and neighbors. Second, this age range is important in the development of both a science mindset and a science identity, where the child explores, experiments, seeks to answer their own questions, and develops observational skills that form the basis for future science learning. Finally, it is clear that children, from an early age, are able to describe the natural world and to explain how they know a fact or an explanation of a phenomena to be true or false; that is to say, they are developing the elements of an epistemology in creating their own theory of knowledge (Driver et al. 1996).

Research shows that children evaluate evidence and test ideas as they play (van Schijndel et al. 2015). The play may involve fantasy and make-believe, but are still forms of engagement that encourage the evaluation and testing of ideas. It is important to view children's ideas of natural phenomena as part of their coherent framework and not as misconceptions or mistakes that need to be corrected. The interpretive models of the natural world that children form may later need further development or even Kuhnian paradigm shifts, where older ideas are replaced by new ones that better explain an observation. The development of scientific knowledge in children is less about the limitations of a child's developmental stage and more about the progression of their own existing ideas and experiences with a phenomenon and how both of these are evolving (Driver 1985).

Since both child development and idea development occurs in each child at a different rate, the program and concepts behind preschool and early primary grade science education apply just as well to all of primary school and beyond. Of key importance is an understanding of the concept of "learning progressions" in astronomy, which describes the process by which a student progresses towards a big idea, such as motion in the solar system. The idea gains sophistication and explanatory power as the child receives educational instruction over several school grade levels and is able to make greater sense of seemingly disparate phenomena (Plummer et al. 2015).

The training of preschool teachers in astronomy and science concepts, and in teaching techniques, is important to the success of these programs. For example, a Greek study of early childhood educators showed that a large fraction of them could not distinguish the differences between astronomy as a science and astrology as a pseudoscience (Kallery 2001). Plummer et al. (2010) showed that an open inquiry experience could positively affect a large percentage of students in a class on elementary school science teaching methods. Their study also highlighted the need for additional instruction such as computer-assisted and planetarium program-based activities for the teachers. Studies have also highlighted the importance of providing high quality "educative curriculum materials" (e.g. Pompea & Walker 2017) for use in the classroom that support teachers in improving their pedagogical content knowledge after they have completed their limited pre-service teacher training in this area. A strong case can be made for improving the competence and confidence of pre-service teachers of young learners as made by a New Zealand study that indicates that science subject knowledge is one of the pre-service teachers' weakest competencies. However, these same pre-service teachers do not appreciate that they are inadequately prepared to teach science (Garbett 2003). The *My Sky Tonight Program* has produced excellent educational materials for preschool education and even for planetariums (Plummer et al. 2016; ASP 2019) and is highly recommended for parents and teachers of preschool children.

## 5.13    Projects with Research Data

Astronomy is one of the sciences leading the Big Data field (Dillon, 2015). Astronomy continues to acquire large amounts of data, more rapidly than the rate at which it can be analyzed. In addition, regarding terms of data openness, most astronomical datasets are now freely and publicly available, which is a great opportunity for astronomy education. Nevertheless, students must not be merely passive recipients of data-based facts or concepts. Students (and teachers) need to become data literate, including planning, acquiring, managing, analyzing, and inferring from data (Ferguson 2017). Astronomy education should — like astronomy research — use data to describe the universe and answer questions with the help of data analysis tools and visualizations. In other words, today's students must learn to work and think with data at an early age, so they are prepared for the data-driven society in which they live.

Engaging with real research, data also has the potential to bring students close to the core research enterprise. Several astronomical organizations have used astronomical data in educational activities and programs. Notably, the project ESO/ESA Astronomy Exercise Series (Bacher & Christensen 2001) included observations from the NASA/ESA Hubble Space Telescope and the ESO telescopes, such as measuring the distance to Supernova 1987A, measuring the distance to M100 with the help of Cepheid variable stars, and determining the mass of the black hole at our Milky Way's center. More recently, the educational project "Black Holes in My School" (Golabz 2019), which also uses robotic telescopes, is an example of using an inquiry-based learning approach to explore stellar black hole candidates. "Big Data" Teen Astronomy Cafés (Pompea & Walker 2018) have been designing and testing best practices with high school students to promote an understanding of modern astronomy research with its emphasis on large data sets, data tools, and visualization tools. In fact, the educational program of the LSST will utilize the voluminous LSST research data sets in its classroom research projects (Jacoby 2013; Bauer 2019).

## 5.14    Citizen-Science and Gamification Projects

Citizen science is now extremely popular in terms of having interesting astronomical topics (Marshal et al. 2015) and the overall involvement of the participants (Radick et al. 2010, Land-Zandstra 2016). The experimental and participatory aspects of citizen science make this approach potentially relevant to education, and several authors (Hiller & Kitsantas 2014; Vitone et al. 2016) have identified this prospective role of citizen science in astronomy education. However, even today, there are no reference projects or activities with learning approaches fully integrated into citizen science projects. Some initial attempts for more direct involvement of citizen science in astronomy education were abandoned (e.g. Zoo Teach: http://www.zooteach.org/). On the other hand, the application of game elements and principles in astronomy educational contexts (also known as gamification) have been positive. For example, some participants in the computer game Cerberus — which allows players to tag surface features on Mars — were able to assimilate deeper concepts about Martian geology, and the data analyzed by these players were of high quality and could be used to support scientific research (van 't Woud 2012).

## 5.15    Robotic Telescope-based Projects

The nature of astronomy often has made it difficult to provide education stakeholders, like teachers and students, access to cutting-edge telescopes. However, robotic telescopes have been used for education purposes now over several decades, with success in many areas, such as in the teaching of STEM skills (McLinn 2011). In their review, Gomez and Fitzgerald (2017) identified the primary reason for the success of telescopes in learning environments, which is providing students a real experience to engage with astronomy's methods and techniques, which involve observation, simulation, and theory.

In short, robotic telescope programs provide real scientific data of astronomical objects that usually cannot be seen with the naked eye, offering strong links with STEM fields, like physics, and utilizing concepts dealing with light, gravity, and instrumentation. Gomez and Fitzgerald (2017) have also identified some challenges for robotic telescope educational program development, including the "quality of available supporting material for robotic telescope usage, reliability and scalability of telescope resources, instructor barriers, image and information technology issues for educational use, and lack of effective evaluation within literature."

## 5.16    Evaluation of Informal Science Programs

The evaluation of informal science education programs is an important step in understanding the efficacy of educational programs. Informal science education project evaluation takes many forms, and an excellent framework for summative evaluation is available (Allen et al. 2008), as is a hand-book for project evaluation in general (Stevens et al. 1993). For project leaders of informal STEM projects, a very useful and practical resource for managing evaluations is by Bonney et al. (2016). Several kinds of evaluation may be necessary, and the different types are now defined along with their value in different project phases, adapting from Allen et al. (2008).

"Front end evaluation" is used to determine the knowledge level and interests of your audience and can help you develop projects that meet the needs and interests of your audience. More importantly, it can keep you from developing projects of little interest that have little need or demand for them. This type of evaluation may be especially valuable in museums; an excellent introduction for museums is given by Dierking and Pollack (1998) and for research initiatives, by Storksdieck and Falk (2004).

"Formative evaluation" is critical to determining which educational strategies are working or not working in a project, so adjustments can be made. It is an iterative testing process with your target audience and is designed to allow feedback and corrections early in a project when improvements can still be made. Some projects require additional evaluation near the end, and use "remedial evaluation" to tweak the nearly final or finished products in order to improve the project's "deliverables."

"Summative evaluation" measures the impact of a "settled" project and is often a requirement from the funder of larger projects. It can be used by the funder to justify the project and assess whether the project has produced the appropriate results and value for its expense, which is often public money. The assessment is linked to the project's goals and objectives.

A project should have as its goal the creation of at least a formative and summative evaluation plan. These plans are extremely helpful in encouraging the project organizers and designers to clarify their goals and objectives and to produce a "logic model" of how they want to achieve their objectives. The logic model is helpful in explicitly stating how you think the project will work. It can also be used to assess risk since certain linkages in the project will appear to be tentative or unworkable, once they have been committed to paper. If the logic model has poor logic or an incomplete flow, this will become apparent after some reflection.

The implementation of a "backwards design" approach can also assist the project designers in understanding what they hope to achieve, the best way to achieve it, and an understanding of how they will know when they have achieved it (Wiggins & McTighe 2005).

## 6. SHIFTING THE ASTRONOMY EDUCATION PARADIGM

The astronomy education ecosystem and the educational niches that astronomers inhabit have been the focus of this review. Even with technological advances such as robotic telescopes and computers, the educational system is remarkably resistant to change. The attitude that learners are deficient, empty vessels to be filled by educators still persists. However, there are a number of transformative efforts in how astronomy education projects are designed and executed that have potential to transform the experience of astronomy learners as well as the astronomy education ecosystem. We describe a few of them here.

### 6.1 Design Thinking Approaches

Design thinking approaches in education have been identified as meaningful ways to integrate the different STEAM disciplines (Moore et al. 2014), including positive attitudes towards science and technology careers (Vossen et al. 2018) and skill-development, such as problem-solving, creativity, communication, and teamwork (Guzey et al. 2016). , The design thinking approach usually follows research and design techniques – essential for astronomy: discovery of the challenge, interpretation, idea development (also called ideation), experimentation, and development (what's next?) (See **Figure 4** based on Fierst et al. 2011).

Astronomy educators have been employing versions of these models in their astronomy activities; Open Astronomy Schools (2019) and Dark Sky Rangers (Walker 2012; Nunes & Dourado 2017) are two examples of educational projects using this approach.

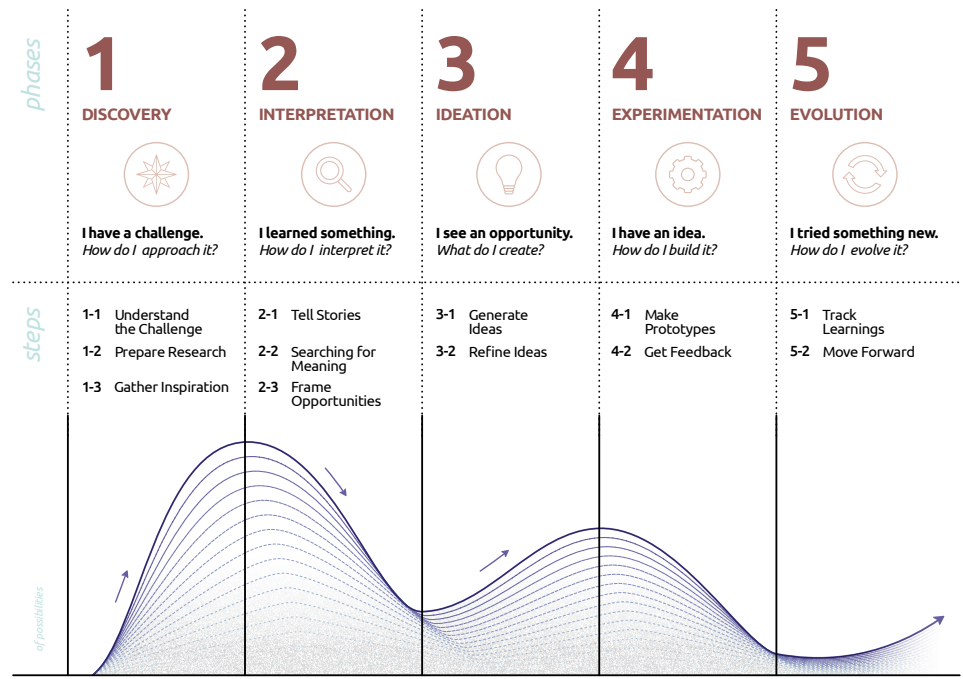

Figure 4

Step-by-step research & design thinking approach. Adapted from NGSS (2012 and 2013) and Fierst et al (2011)

## 6.2 Large-Scale Educational Programs

Because of our interconnected world, there are many programs implemented on a global scale. These large-scale educational programs are designed to enable participation of large numbers of geographically dispersed people in a program with specific objectives. Several large-scale educational programs related to astronomy have been implemented in the last decades, namely the World Year of Physics 2005 (Stone 2004), World Space Week (UN 2019) annual since 1999, the International Year of Astronomy 2009 (Russo & Christensen 2010; Pompea et al. 2010; Hesser et al. 2010), the International Year of Light 2015 (and consequent International Day of Light since 2016) (Dudley et al. 2016), and more recently, the centenary celebrations of the International Astronomical Union. All these initiatives had specific educational goals.

For instance, the International Year of Astronomy 2009 intended to "support and improve formal and informal science education in schools as well as through science centers, planetariums and museums [by] developing formal and informal educational material and distributing all over the world" (Russo & Christensen 2010). The International Day of Light's goal was to "build worldwide educational capacity through activities targeted on science for young people, addressing issues of gender balance, and focusing especially on developing countries and emerging economies" (Dudley et al. 2016). Furthermore, the focus of World Space Week is "providing teachers with classroom activities that use space to excite students about learning" (UN 2019). Also, the International Astronomy Union's purpose is to "support and improve the use of astronomy as a tool for education" (Rivero-Gonzalez et al. 2018).

All these large-scale programs have some common features, discussed in detail in Russo (2015), such as that large-scale educational programs should be based on strong and relevant science cases. Equally important, they should engage with many stakeholders, not only in science, academia, and governance, but also in less traditional communities like artistic groups. The organizers implementing the projects at the local level need to participate in the decision-making process as well as leadership, providing a certain level of ownership, pride, and importance toward the implementation of the programs, with a global coordination body to ensure that this happens. Planning (including impact assessment) should start as soon as the concept is developed. From the start, experts in impact assessments of science education programs should be incorporated into the global coordination team and provide input on all aspects of implementation.

## 6.3    Upstream Engagement with Educational Stakeholders

Astronomy's presence in the national educational curricula is irregular, as some countries have astronomy subjects at different levels and others have no astronomy topics. The Space Awareness project (2019) did a preliminary assessment of the European curriculum, and more recently, Salimpour et al. (2019) reviewed the astronomy presence in the Organisation for Economic Co-operation and Development (OECD) countries' curriculum. Salimpur et al. (2019) noted that astronomy-related content was prevalent across the 52 curricula of the European and OECD countries. In terms of content, topics relating to basic astronomy were the most prevalent, while topics related to cosmology and current research in astronomy were not as common. This could be due to the limited content-knowledge of current astronomy topics and issues with curriculum developers, policymakers, and even teachers. Nonetheless, astronomers can and should engage with curriculum developers to bring cutting-edge astronomy research into the national curricula.

Some other good practices to impact the formal educational content encompasses working with educational textbook publishers. Good examples of this include the use of Space Scoop text in English language books (Reynods 2014), South African textbooks, and Dutch Physics textbooks. Astronomers can work closely with textbook publishers to foster the use of more astronomy educational resources in educational material across different school subjects, especially by reviewing and developing the content of specific textbooks. These approaches can provide a much wider reach of astronomy content in the education sector.

## 6.4    Global Citizenship Education: Learning through Perspective

The cosmic perspective that astronomy provides makes it an important tool for global citizenship education. Global Citizenship Education (GSED) is UNESCO's educational approach to tackle some of the global issues that threaten peace (UNESCO 2014) and sustainability (climate change, human rights violations, inequality, and poverty). Global Citizenship Education empowers learners of all ages to assume active roles, both locally and globally, in building more peaceful, tolerant, inclusive, and secure societies. The Pale Blue Dot concept, as presented by Carl Sagan, gave us the astronomy perspective for the three domains of learning for the UNESCO's Global Citizenship Education program (UNESCO 2014):

— **Cognitive:** „I wanted to be a scientist from my earliest school days. The crystallizing moment came when I first caught on that stars are mighty suns, and how staggeringly far away they must be to appear to us as mere points of light" (Sagan 1994a).

— **Socio-emotional:** „Fanatical ethnic or national chauvinisms are difficult to maintain when we see our planet as a fragile blue crescent fading to become an inconspicuous point of light against the bastion and citadel of the stars" (Sagan 1980).

— **Behavioral:** "Look again at that dot. That's here. That's home. That's us. It makes clear our responsibility to deal more kindly with one another, and to preserve and appreciate the pale blue dot" (Sagan 1994b).

Several educational programs have been using this approach, including Universe Awareness (UNAWE) and Big History.

Universe Awareness is a global science education program using the beauty and grandeur of the universe to encourage young children (4 to 10 years old), particularly those from an underprivileged background, to have an interest in science and technology and foster their sense of global citizenship. There is evidence that using astronomy intervention contributes to the motivation, science knowledge, science skills development, and intercultural awareness of learners (Kimble 2013).

Big History is an educational and storytelling approach utilizing the "history of the universe over time through a diverse range of disciplines that spans physics, chemistry, biology, anthropology, and archaeology, thereby reconciling traditional human history with environmental geography and natural history" (Simon et al. 2015). In the last five years, this approach has received more attention and funding (Sorking 2014) as well as a complete teaching website. The Big History Project was developed with content focusing on evidence showing that students using this approach have some gains in reading, writing, and content knowledge (Big History Project 2015).

## 6.5    Beyond Science Literacy: Science Capital

The complexity and diversity of methodologies and interaction approaches of individuals related to astronomy is rapidly changing. There are new challenges in the methods of designing, implementing, and assessing astronomy education, which have large implications in the construction and measurement of science literacy of individuals (and, of course, students). Authors (Bucchi & Trench 2014) have asked for a move from the simple facts approach to literacy to a capital approach. Science capital, a concept introduced by a UK policy report — Aspires (Archer 2013) — is the sum of all science-related knowledge, attitudes, experiences, and resources that individuals build up throughout their life. This includes what science they learn and know from all different types of learning environments (from classrooms to out-of-school programs), what they think about science, the people they know who understand science, and the day-to-day engagement they have with science (Archer et al. 2015). This capital is closely connected with knowledge but also with cultural and social capital (institutionalized and/or embodied through knowledge, media consumption, and social networks).

It is important that lifelong diverse learning opportunities are available for children, adolescents, and adults within a robust community-wide system of opportunities for science education (Falk et al. 2015). Having this perspective of an educational ecosystem creates an emphasis on building partnerships that can support and follow the STEM learning of young people across multiple settings (Schatz & Dierking 1998).

Astronomy education must also embrace this approach and support individuals' development of astronomy capital. Indeed, astronomy education programs must find ways for astronomers, students, and teachers to collaborate. For astronomy to be relevant to students' lives and citizenship, astronomy education should develop opportunities for learners to participate at unique stages of the astronomy enterprise, including collecting and analyzing astronomy research data (Roth & Barton 2004).

An understanding of "science capital" offers the practitioner or project designer a powerful conceptual lens to aid in identifying interventions which can address disparities in science engagement and participation (DeWitt et al. 2016). Cultivating a "growth mindset" (Dweck 2016) in adolescents is also an important element that can aid science achievement and a more robust approach to science. A growth mindset emphasizes that a person can improve their skill set rather than remaining self-critical and fearing failure, or at least what is perceived as failure. Encouraging a growth mindset has been effective in art-science programs that emphasize a creative cycle of design experimentation, prototyping, evaluation, and redesign (Conner et al. 2017; Tsurusaki et al. 2017). Recent research using online interventions that teach students that intellectual abilities can be developed (i.e. a growth mindset) have proven effective in improving grades in lower-achieving students, even though the intervention lasted less than one hour (Yeager et al. 2019).

## 6.6    Astronomy as an Open-Schooling Approach

Due to its links with diverse subjects (Miley 2009), astronomy has the major potential to support the demanded shift towards Open Schooling (Hazelkorn 2015) approaches by linking astronomy with other subjects, disciplines, and industries. These connections would help show how interdisciplinarity can contribute to our understanding and knowledge of scientific principles and solve societal challenges. Such a linking would also strengthen the connections and synergies between science, creativity, entrepreneurship, and innovation, which might support astronomy graduates in acquiring key competencies to ease the transition from research to non-research jobs.

Astronomy is a natural science to foster collaborations between formal, non-formal, informal educational providers, enterprises, and civil society — the so-called Open Schooling. Astronomy naturally engages with different stakeholders, such as professionals from space industries and enterprises, civil entities, and wider society. Moreover, astronomy education must actively involve stakeholders in bringing real-life projects into educational programs. As a result, key partnerships will be created between teachers, students, researchers, innovators, professionals in enterprise, and other stakeholders in science-related fields, in order to work on real-life challenges and innovations, including associated ethical, social, and economic issues. Such partnerships will foster networking as well as distributing astronomy and technology research findings amongst teachers, researchers, and professionals across different enterprises (e.g. start-ups, small and medium sized enterprises, and large corporations).

## 6.7    Art and Visual Thinking Approaches

Astronomy education is now appreciating the power of the art-based approaches it has used for decades. The role of the arts in learning processes has been extensively studied, and there is strong evidence that, through art activities, students' general attitude towards learning and school can improve (Gardiner 1996). Moreover, learning art skills forces mental "stretching" which is useful to astronomy education, such as spatial visualization (Plumer 2014; Cole et al. 2018). Art has been used most often in primary education, such as in the Universe Awareness' guide (in Spanish) on Art & Astronomy Activities (Henao 2015), which provides several activities combining astronomy and different art techniques. Combining art with astronomy education programs also provides powerful contexts for equity-focused teaching and learning through efforts to leverage young people's interests and cultural resources (Art + Science, 2016). The inspiration of astronomical phenomena (White 2000) and the presence of artist-in-residence programs at observatories have helped astronomers appreciate the alternative perspectives that art can bring.

Of great current interest is that art approaches to STEM may be key to broadening participation in STEM learning. The spatial abilities of adolescents are an attribute that correlates with occupations in STEM fields in an 11-year study of 400,000 students (Wai et al. 2009). However, this effect is noticeable only for males and research has focused on identifying barriers to entering STEM fields for females with similar abilities (Tzou et al. 2014). The perception by girls that science is uncreative and passionless is one area of concern (Miller et al. 2006) that must be addressed.

Research suggests that girls who gravitate toward art often have strong visual-spatial abilities that would serve them well in science careers. Visual-spatial thinking skills entails the ability to address the shapes, places, and movements of objects relative to each other. These skills are the ones used in map reading, assembling something from a diagram, or in trying to rearrange the furniture in a room. Visual-spatial thinking skills are especially useful in astronomy learning. For example, they help in understanding concepts of scale and in constructing 3D models of objects such as galaxies and visualizing stellar orbits. These skills provide a solid footing for STEM learning; they can be taught to children of all ages (Newcombe 2010; National Research Council and Geographical Sciences Committee 2005).

One goal of STEAM programs is to connect these girls to science at an age when their larger identity is still developing in order to attract them to STEM careers. A well-designed STEAM program can inspire girls with artistic talents and interests to enter STEM careers. It can do this through an arts-based approach that builds on their existing strengths in visual-spatial reasoning (Conner et al. 2017; Tzou et al. 2014; Carsten Conner et al. 2019).

## 6.8    Indigenous Education

Many astronomical facilities have the privilege of being located on the lands of indigenous peoples; these mountain tops are notable for their pristine air and dark skies. These include the observatories on Maunakea (Hawai'i), dzil nchaa si'an (Mt. Graham Arizona), and Iolkam Du'ag (Kitt Peak Arizona). As observatories strive to be good neighbors and valued community partners to indigenous groups, there is much to be learned in order to have productive educational collaborations with Indigenous Nations such as the Tohono O'odham Nation, on whose land Kitt Peak National Observatory is located. Indigenous people are severely and chronically underrepresented in the STEM fields, when compared to their representation in the total population. To address this issue in collaboration programs between observatories and indigenous groups, half measures will not be useful.

To understand the context of working in partnership with tribal Nations, U.S. astronomers might begin with a brief study of their nation's and region's history which illustrates dramatically how the government and educational institutions have consistently betrayed the trust of indigenous people on numerous occasions. Educational institutions have been particularly damaging to these cultures, through language suppression, mandatory Christian religious instruction, and the use of boarding schools that relocated young children away from their families, homes, and communities (Ferguson 1997).

The context of the role of science in society and how it is taught is also relevant. An understanding of both the convergence and divergence of Western and Native perspectives on science, as well as STEM learning practices, is critical for collaborations. For example, the discord between the cultural component of learning science in school and the Native American cultural ways of knowing can produce a sense of alienation and disengagement on the part of students, leading to poor school performance. In particular, Native American students desire broader contexts for their science knowledge which can create connections and relevance of this knowledge to their family, community, tribal values, and their tribal practices (Bang et al. 2010). Collaborative projects must emphasize community-based design and rely on efforts to create the comprehensive participation of community members.

Cajete (1994, 1999) describes in depth the creative possibilities of an ecology of indigenous education based on cultural orientations to learning and teaching. This approach includes both a much more systems-oriented approach and an emphasis on personal growth. These values are more prominent than is usually found in traditional educational approaches. By applying this perspective, the result is a more flexible, viable, and effective approach for program creation.

Peticolas et al. (2012) describes the cultural disconnections between Western scientists/educators and Native communities in their worldviews and ways of knowing. In the U.S. *Cosmic Serpent: Bridging Native Ways of Knowing and Western Science in Museum Settings* project, workshops were conducted by the Indigenous Education Institute and the University of California, Berkeley Space Sciences Lab. The series of workshops brought together 115 science education practitioners from 19 tribal museums and 41 science/natural history and cultural museums with 23 tribal community members and 24 "bridge people" with knowledge of both Indigenous and Western science. These workshops yielded a number of lessons on how to start genuinely collaborative cross-cultural projects that reflect both world views. A key conclusion was that all stakeholders must pursue "collaboration with integrity". Resources from the project are extensive, available, and useful for both astronomers and educators (Cosmic Serpent 2019).

Other programs that incorporate these lessons include the program *Indigenous Worldviews in Informal Science Education* (Venkatesan & Burgasser 2017; Hawkins & Swimmer 2016). These programs encourage stronger collaboration of the education efforts of observatory with their indigenous community educational partners. We heartily endorse this perspective.

## 7.    CONCLUDING REMARKS

Evidence-based astronomy education is a field that is evolving rapidly in breadth, depth, and sophistication. There are many diverse and challenging roles for astronomers. The occasional or dedicated practitioners in our astronomy education field now include nearly every astronomer worldwide who values education and the role that their own teachers and educators have played in guiding them.

As the field has evolved, and with much new research available, a new level of professionalism and intentionality has framed our activities. We now debate the value of the educational parameter spaces and niches we want to explore, much as astronomers discuss the wavelength ranges, spatial resolutions, spectroscopic dispersions, and observing cadences of a new instrument. We consider risks, costs, and throughput; educational etendue is appraised as we design the intensity and reach of our programs. In a field teeming with experience and expertise, the universal experience of attending school may be understood more for its ability to create misconceptions than as an educational qualification. The lack of complaints and the polite applause of our educational audiences when we "outreach" to them now means little in evaluating our performance and effectiveness, given the wide availability of evaluation techniques and instruments. Rather, if we are satisfied with their lukewarm responses, we are only demonstrating that we are lackadaisical in understanding our education partners and their self-effacing manner.

The experienced EPE practitioner often functions like a combination project and systems engineer by integrating these two roles and skill sets to make sure that the intention and end goals of educational projects are preserved and achieved. We do this by preventing novice mistakes, identifying how the project pieces and stakeholders fit together, designing effective interfaces, conducting the needed testing, and solving pressing problems, often by adding additional expertise from outside the project.

Our review was meant as a guide to the relevant education research for astronomers who are involved with, or want to get involved with, the formal and informal education ecosystem. We wanted to help astronomers efficiently design effective programs that work with students and families, schools, teachers, visitor centers, planetariums, museums, teacher and student research programs, instructional materials development, and professional development programs. We hope this review has encouraged your professional growth in astronomy education. If it has helped you frame your educational systems thinking and project thinking around these evidence-based approaches, then we count our education efforts in this review as a success.

## ACKNOWLEDGEMENTS

This work was supported by the National Optical Astronomy Observatory, which is operated by the Association of Universities for Research in Astronomy (AURA), Inc. under cooperative agreement with the U. S. National Science Foundation.

Stephen Pompea acknowledges the University of Arizona where he is an Affiliate Professor at Steward Observatory and an Adjunct Professor at the College of Optical Sciences. Dr. Pompea also acknowledges the support of Leiden Observatory and Leiden University, where he is a visiting Professor.

Russo Russo acknowledges the support of Leiden Observatory and Department of Science Communication & Society of the Leiden University where he is a University Professor in Astronomy & Society and leads the Astronomy & Society Group. Dr. Russo was the global coordinator for the United Nations' International Year of Astronomy 2009. He is also affiliated with the Institute of Space Sciences in Portugal and the Institute of Public Communication of Science and Technology in Brazil. Dr. Russo was the founder and first editor-in-chief of Communicating Astronomy with the Public Journal. His work has received several awards, such as Seeds Special Award 2009,

Scientix Best Educational Resource in 2015 and 2016, Most Innovative Educational Activities in 2017 and 2018 by HundrED.

We wish to acknowledge for their expertise and assistance: Hidehiko Agata, John Akey, Tomita Akihiko, Doug Arion, Warren Barta, Peter Barthel, Tobias Beuchert, Paulo Bretones, Bob Blum, Craig Blurton, Roger Culver, Kelvin Day, Urban Eriksson, Rick Fienberg, Michael Fitzgerald, Andrew Fraknoi, Duane Garnett, Edward Gomez, Alan Gould, Jorge Rivero Gonzalez, Michelle Hall, Isabel Hawkins, Chris Impey, Kevin McLin, George Miley, Tibisay Sankatsing Nava, Carmen Pantoja, Donald Pegler, Laura Peticolas, Derrick Pitts, Nancy Regens, Gustavos Rojas, Dennis Schatz, Dave Silva, Malcolm Smith, Tim Spuck, Martin Storksdieck, Wayne Sukow, David Ulmer, George Wallace, and Vivian White., Steve also thanks the current and former members of the NOAO education team North and South. Pedro would like to thank to the present and past members of Leiden Observatory's Astronomy & Society group and the Leiden University's Department of Science Communication & Society.